\documentclass[sigconf, natbib, screen=true, anonymous=false, review=false]{acmart}

\acmSubmissionID{1061}

\usepackage{mathtools}
\usepackage{fixmath}
\usepackage{graphicx} %
\usepackage{subfigure}
\usepackage{xcolor}
\usepackage{siunitx}
\usepackage[inline]{enumitem}
\usepackage{geometry}
\usepackage{booktabs}
\usepackage{multirow}
\usepackage{acronym}
\usepackage[skip=2pt]{caption}

\AtBeginDocument{%
  \providecommand\BibTeX{{%
    \normalfont B\kern-0.5em{\scshape i\kern-0.25em b}\kern-0.8em\TeX}}}

\renewcommand{\vec}[1]{\mathbold{#1}}
\newcommand{\mat}[1]{\mathbold{#1}}
\newcommand{\tensor}[1]{\mathbf{#1}}
\newcommand{\ve}{\vec{e}} %
\newcommand{\vh}{\vec{h}} %
\newcommand{\vd}{\vec{d}} %
\newcommand{\vq}{\vec{q}} %
\newcommand{\mE}{\mat{E}} %
\newcommand{\mA}{\mat{A}} %
\newcommand{\mQ}{\mat{Q}} %
\newcommand{\mD}{\mat{D}} %
\newcommand{\mW}{\mat{W}} %
\newcommand{\mR}{\mat{R}} %
\newcommand{\mc}[1]{\mathcal{#1}}

\newcommand{\V}{\mathcal{V}}
\newcommand{\nV}{|\mathcal{V}|} %
\newcommand{\R}{\mathbb{R}}

\newcommand{\headernodot}[1]{\vspace*{1mm}\noindent\textbf{#1}}
\newcommand{\header}[1]{\headernodot{#1.}}

\acrodef{DR}{dense retrieval}
\acrodef{GR}{generative retrieval}
\acrodef{MVDR}{multi-vector dense retrieval}
\acrodef{PAWA}{prefix-aware weight-adaptive}
\acrodef{NP}{non-parametric}

\copyrightyear{2024}
\acmYear{2024}
\setcopyright{acmlicensed}
\acmConference[SIGIR '24] {Proceedings of the 47th International ACM SIGIR Conference on Research and Development in Information Retrieval}{July 14--18, 2024}{Washington, DC, USA.}
\acmBooktitle{Proceedings of the 47th International ACM SIGIR Conference on Research and Development in Information Retrieval (SIGIR '24), July 14--18, 2024, Washington, DC, USA}
\acmISBN{979-8-4007-0431-4/24/07}
\acmDOI{10.1145/XXXXXX.XXXXXX}

\begin{document}

\title{Generative Retrieval as Multi-Vector Dense Retrieval}

\author{Shiguang Wu}
\orcid{0000-0002-4597-5851}
\affiliation{%
  \institution{Shandong University}
  \streetaddress{72 Binhai Road, Jimo}
  \city{Qingdao}
  \country{China}
  \postcode{266237}
}
\email{shiguang.wu@mail.sdu.edu.cn}

\author{Wenda Wei}
\orcid{0009-0007-3504-0933}
\affiliation{%
  \institution{Shandong University}
  \streetaddress{72 Binhai Road, Jimo}
  \city{Qingdao}
  \country{China}
  \postcode{266237}
}
\email{weiwenda@mail.sdu.edu.cn}

\author{Mengqi Zhang}
\orcid{0000-0001-6831-0740}
\affiliation{%
  \institution{Shandong University}
  \streetaddress{72 Binhai Road, Jimo}
  \city{Qingdao}
  \country{China}
  \postcode{266237}
}
\email{mengqi.zhang@sdu.edu.cn}

\author{Zhumin Chen}
\orcid{0000-0003-4592-4074}
\affiliation{%
  \institution{Shandong University}
  \streetaddress{72 Binhai Road, Jimo}
  \city{Qingdao}
  \country{China}
  \postcode{266237}
}
\email{chenzhumin@sdu.edu.cn}

\author{Jun Ma}
\orcid{0000-0003-0203-4610}
\affiliation{%
  \institution{Shandong University}
  \streetaddress{72 Binhai Road, Jimo}
  \city{Qingdao}
  \country{China}
  \postcode{266237}
}
\email{majun@sdu.edu.cn}

\author{Zhaochun Ren}
\orcid{0000-0002-9076-6565}
\affiliation{%
  \institution{Leiden University}
  \streetaddress{Snellius, Niels Bohrweg 1}
  \city{Leiden}
  \country{The Netherlands}
  \postcode{2333 CA}
}
\email{z.ren@liacs.leidenuniv.nl}

\author{Maarten de Rijke}
\orcid{0000-0002-1086-0202}
\affiliation{%
  \institution{University of Amsterdam}
  \streetaddress{Science Park 900}
  \city{Amsterdam}
  \country{The Netherlands}
  \postcode{1098 XH}
}
\email{m.derijke@uva.nl}

\author{Pengjie Ren}
\orcid{0000-0003-2964-6422}
\authornote{Corresponding author.}
\affiliation{%
  \institution{Shandong University}
  \streetaddress{72 Binhai Road, Jimo}
  \city{Qingdao}
  \country{China}
  \postcode{266237}
}
\email{jay.ren@outlook.com}

\renewcommand{\shortauthors}{Shiguang Wu et al.}

\begin{abstract}
    \Acl{GR} generates identifiers of relevant documents in an end-to-end manner using a sequence-to-sequence architecture for a given query.
    The relation between \acl{GR} and other retrieval methods, especially those based on matching within dense retrieval models, is not yet fully comprehended.
    Prior work has demonstrated that \acl{GR} with atomic identifiers is equivalent to single-vector dense retrieval.
    Accordingly, \acl{GR} exhibits behavior analogous to hierarchical search within a tree index in dense retrieval when using hierarchical semantic identifiers. 
    However, prior work focuses solely on the retrieval stage without considering the deep interactions within the decoder of \acl{GR}.
    
    In this paper, we fill this gap by demonstrating that \acl{GR} and \acl{MVDR} share the same framework for measuring the relevance to a query of a document.
    Specifically, we examine the attention layer and prediction head of \acl{GR}, revealing that \acl{GR} can be understood as a special case of \acl{MVDR}.
    Both methods compute relevance as a sum of products of query and document vectors and an alignment matrix.
    We then explore how \acl{GR} applies this framework, employing distinct strategies for computing document token vectors and the alignment matrix.
    We have conducted experiments to verify our conclusions and show that both paradigms exhibit commonalities of term matching in their alignment matrix.

    Our finding applies to many \acl{GR} identifier designs and provides possible explanations on how \acl{GR} can express query-document relevance.
    As \acl{MVDR} is the state-of-the-art dense retrieval method currently, understanding the connection between \acl{GR} and \acl{MVDR} is crucial for shedding light on the underlying mechanisms of \acl{GR} and for developing, and understanding the potential of, new retrieval models. %
\end{abstract}

\begin{CCSXML}
<ccs2012>
   <concept>
       <concept_id>10002951.10003317.10003338</concept_id>
       <concept_desc>Information systems~Retrieval models and ranking</concept_desc>
       <concept_significance>500</concept_significance>
       </concept>
 </ccs2012>
\end{CCSXML}

\ccsdesc[500]{Information systems~Retrieval models and ranking}

\keywords{Generative retrieval, Dense retrieval, Multi-vector dense retrieval}

\maketitle

\acresetall

\section{Introduction}
\label{sec:intro}

In recent years, the advent of pre-trained language models has catalyzed the popularity of neural-based retrieval models within the information retrieval community~\citep{karpukhinDensePassageRetrieval2020,niLargeDualEncoders2021,khattabColBERTEfficientEffective2020,qianMultiVectorRetrievalSparse2022,tayTransformerMemoryDifferentiable2022a}.

\header{Neural-based retrieval models}
\Ac{DR}, as one of the effective neural-based retrieval methods, has achieved the state-of-the-art ranking performance on multiple benchmarks~\citep{karpukhinDensePassageRetrieval2020,niLargeDualEncoders2021,khattabColBERTEfficientEffective2020}.
Several approaches have been proposed to use multiple vectors to represent documents or queries, a.k.a., \acfi{MVDR}~\citep{khattabColBERTEfficientEffective2020,qianMultiVectorRetrievalSparse2022,zhouMultiVectorAttentionModels2021}.

Recently, \acfi{GR} has emerged as a new paradigm in information retrieval.
It aims to generate identifiers of relevant documents for a given query directly and parametrizes the indexing, retrieval, and ranking process in dense retrieval systems into a single model.
\Ac{GR} adopts a sequence-to-sequence architecture model and is trained to directly map queries to their relevant document identifiers.

\header{\Acl{GR} vs. \acl{DR}}
Dense retrieval models typically employ encoders, e.g., BERT~\citep{devlinBERTPretrainingDeep2018}, for encoding both queries and documents, while the \acl{GR} model adopts an encoder for query encoding and a decoder for identifier generation.
Despite their superficial differences, \acl{DR} and \acl{GR} share key characteristics in their query-document relevance computations.
When the two methods use document identifiers such as sub-strings, titles, or semantic IDs as representations for documents, relevance to a query of a document is computed as the dot product of two vectors in both methods.
Dense retrieval involves using the direct product of the query vectors and document vectors as the relevance, while \acl{GR} leverages the product of the last latent state from the decoder at each position with the prediction head, a.k.a., the word embedding lookup table.
Consequently, a natural question that arises in this context:
\begin{quote}
    \textit{How is \acl{GR} related to \acl{DR}?}
\end{quote}
Although \ac{GR} has shown promising results in various benchmarks as a new end-to-end retrieval paradigm~\citep{liLearningRankGenerative2023,wangNeuralCorpusIndexer2023,sunLearningTokenizeGenerative2023,yangAutoSearchIndexer2023,leeGLENGenerativeRetrieval2023a}, relatively few publications have closely examined how \ac{GR} models work.
\citet{nguyen-2023-generative} have shown that \ac{GR} using atomic identifiers can be viewed as a variant of bi-encoders for dense retrieval because the word embedding lookup table in \acl{GR} works exactly the same as the flat index in dense retrieval.
Thus, we can partially respond to the above question that \ac{GR} with atomic identifiers \emph{is} single-vector dense retrieval.
Although atomic identifiers are considered non-mainstream in \ac{GR}, it offers an insightful perspective on the matter.
\citet{nguyen-2023-generative} also discuss that \ac{GR} with hierarchical semantic identifiers exhibits behavior similar to hierarchical search within a tree index in dense retrieval.
However, their discussion focuses only on the retrieval stage without rigorously considering deep interactions within the decoder.

\header{\Acl{GR} as \acl{MVDR}}
In this work, we connect \acl{GR} to a state-of-the-art dense retrieval method, \acl{MVDR}, in a rigorous way.
We illustrate that these two methods exhibit \emph{commonalities in their training targets and a shared focus on semantic matching}.
We first examine the attention layer and the prediction head of \ac{GR} and show that the logits in the loss function can be reformulated to a product of document word embeddings, query token vectors, and attention matrix in Section~\ref{sec:brief_derivation}.
This corresponds to the unified \ac{MVDR} framework introduced in~\citep{qianMultiVectorRetrievalSparse2022,leeRethinkingRoleToken2023}.
In Section \ref{sec:main_comparison} we explore the distinct document encoding and alignment strategy in \ac{GR}.
Specifically, our discussion includes
\begin{enumerate*}[label=(\roman*)]
    \item its simple document encoding and how \ac{PAWA} decoding~\citep{wangNeuralCorpusIndexer2023} and \ac{NP}-decoding~\citep{leeNonparametricDecodingGenerative2023} apply to our framework (Section~\ref{sec:compari_document_encoding}), and
    \item the distinct alignment strategy employed by \ac{GR} compared to \ac{MVDR} and its properties (Section~\ref{sec:compari_alignment_strategy}).
\end{enumerate*}

Our discovery provides reliable explanations of how \ac{GR} can express query-document relevance.
By explaining how the \ac{GR} method models the query-document relevance, we can further understand how \ac{GR} is fundamentally different from dense retrieval methods and adds to the spectrum of neural-based retrieval models.
The connection we present provides the variants of \ac{GR} methods with a theoretical foundation for further improvement.

\header{Contributions}
Our main contributions in this paper are:
\begin{enumerate}[leftmargin=*,nosep]
    \item We offer new insights into \ac{GR} from the perspective of \ac{MVDR} by showing that these methods share the same framework for measuring query-document relevance.
    \item We explore how \ac{GR} applies this framework, employing distinct strategies for document encoding and the alignment matrix.
    \item We also conduct extensive analytical experiments based on the framework to verify our conclusions and illustrate the term-matching phenomenon and properties of different alignment directions in both paradigms.
\end{enumerate}

\section{Related Work}

\headernodot{\Acf{MVDR}} can be seen as a generalization of single-vector dual encoder models~\citep{karpukhinDensePassageRetrieval2020,khattabColBERTEfficientEffective2020}.
Instead of encoding the complete content of both query and documents into a single low-dimensional vector, \ac{MVDR} uses fine-grained token-level modeling for scoring.
Existing \ac{MVDR} models such as ColBERT~\citep{khattabColBERTEfficientEffective2020} compute query-document relevance by selecting the highest-scoring document token for each query token and aggregating the scores.
The postponed token-level interactions allow us to efficiently apply the model for retrieval, benefiting the effectiveness of modeling fine-grained interactions.
\Ac{MVDR} overcomes the limited expressivity of single-vector retrieval and achieves significantly better results across various benchmarks~\citep{karpukhinDensePassageRetrieval2020,niLargeDualEncoders2021,khattabColBERTEfficientEffective2020,qianMultiVectorRetrievalSparse2022,leeRethinkingRoleToken2023,liSLIMSparsifiedLate2023}.
However, due to the cost of storing vectors for each document token, it is challenging to scale the approach to large collections~\citep{khattabColBERTEfficientEffective2020,formalSPLADESparseLexical2021,qianMultiVectorRetrievalSparse2022,leeRethinkingRoleToken2023}.

\headernodot{\Acf{GR}} is an emerging paradigm in information retrieval~\citep{tayTransformerMemoryDifferentiable2022a,zengScalableEffectiveGenerative2023b,wangNeuralCorpusIndexer2023,leeGLENGenerativeRetrieval2023a,pradeepHowDoesGenerative2023,rajput2023recommender}.
It leverages generative models to directly generate identifiers of relevant documents.
This approach originated with \citep{decaoAutoregressiveEntityRetrieval2020,tayTransformerMemoryDifferentiable2022a} and has garnered considerable attention~\citep[see, e.g.,][]{tang-2023-recent}.
Currently, all implementations of the generative retrieval paradigm adhere to an encoder-decoder transformer architecture, e.g., T5~\citep{t52020} and BART~\citep{bart}.
In this method, documents are initially associated with a concise token sequence that serves as an identifier. 
The model is then trained to predict this token sequence autoregressively, using conventional cross-entropy loss.

One notable advantage of the generative retrieval model is its streamlined end-to-end architecture, which requires significantly less disk storage space compared to other retrieval methods.
However, it is important to note that due to the limited supervision of each token, the generative retrieval may not achieve comparable performance when compared to dense retrieval~\citep{pradeepHowDoesGenerative2023,yuan2024generative-memory-burden}.

\header{Connecting dense retrieval and generative retrieval}
\citet{nguyen-2023-generative} show that \ac{GR} with atomic identifiers is equivalent to single-vector dense retrieval.
They compare the inferential processes of  \ac{DR} with a tree index and \ac{GR} with hierarchical identifiers.
However, the former is just an optimized version of the original \ac{DR} without changing the semantic matching method, while the latter also considers the predicted IDs and the query in each generation step, which greatly affects how \ac{GR} would express the relevance, but this is ignored in~\citep{nguyen-2023-generative}.
\citet{yuan2024generative-memory-burden} empirically analyze the error rate at each generation step of \ac{GR} and identify the problem of poor memory accuracy for fine-grained features compared with \ac{DR}.
They straightforwardly integrated \ac{GR} and \ac{DR} into a new coarse-to-fine retrieval paradigm, combining their respective strengths, but circumvented an in-depth discussion of the connection between them.

In this work, we address the limitations listed above by showing that \ac{GR} expresses query-document relevance in the same way as \ac{MVDR}.
This connection is rigorously derived from the decoder of \ac{GR} and can be applied to many identifiers.

\section{Preliminaries}
In this section, we formulate our task and introduce key notation and the mainstream framework of \ac{MVDR} models.

\header{Task definition}
We formulate the retrieval task as ranking by relevance score.
Given a query $q$, we aim to retrieve relevant documents $d$ in $\mc{D}$ by ranking them by their relevance $\operatorname{rel}(d, q)$ to $q$.

\header{Notations}
Table~\ref{table:notation} lists the main notation used in the paper.
We denote the word embedding lookup table in the decoder as $\mE$ and the vocabulary set as $\V$.
Each document $d$ comprises $M$ tokens. %
To ensure uniform length, padding tokens are added or excess tokens are truncated from each document.
The word embedding matrix of $d$ is denoted as $\mE_d \coloneqq [\ve_{d_1}, \ldots, \ve_{d_M}] \in\R^{d\times M}$, and the latent token vector matrix after encoding is $\mD \coloneqq [\vd_1, \ldots, \vd_{M}] \in\R^{d\times M}$.
Each query $q$ with $N$ tokens has token vectors $\mQ\coloneqq [\vq_1, \ldots, \vq_{N}] \in\R^{d\times N}$ after encoding, similar to the documents.

\begin{table}[ht]
\caption{Main notation used in this work.}
\label{table:notation}
\begin{tabular}{ll}
    \toprule
    \textbf{Symbol} & \multicolumn{1}{c}{\textbf{Description}}
    \\
    \midrule
    $\mat{E}$       & word embedding lookup table \\
    $\mE_d$         & document word embedding matrix \\
    $\mD,\mQ$       & document / query token vector matrix \\
    $d_i,q_j$       & document / query tokens \\
    $\vd_i,\vq_j$   & encoded document / query token vector \\
    \bottomrule
\end{tabular}    
\end{table}

\header{Framework for \ac{MVDR}}
\label{sec:MVDR}
Following \citep{qianMultiVectorRetrievalSparse2022,leeRethinkingRoleToken2023}, \ac{MVDR} methods can be represented as a unified framework, in which the relevance between query $q$ and document $d$ is given by:
\begin{equation}
    \operatorname{rel}(d, q)
    = \frac{1}{Z}\operatorname{sum}(\mD^\top \mQ \odot \mA)
    = \frac{1}{Z}\sum_{i, j}\vd_i^\top \vq_j A_{ij},
    \label{eq:framework_mvdr}
\end{equation}
where $\mA$ is the alignment matrix that controls whether a document and query token pair can be matched and contribute to the relevance, and $Z = \sum_{ij}A_{ij}$ is used for normalization and is dropped in many \ac{MVDR} methods.

\header{Alignment strategy} Different \ac{MVDR} models adopt different alignment strategies, and, thus, a different alignment matrix $\mA$.
It is often computed using heuristic algorithms, such as lexical exact match~\citep{gaoCOILRevisitExact2021}, top-$1$ relevant token match~\citep{khattabColBERTEfficientEffective2020}, single-vector alignment~\citep{karpukhinDensePassageRetrieval2020,luanSparseDenseAttentional2021}, or sparse unary salience~\citep{qianMultiVectorRetrievalSparse2022}.

\header{Contrastive loss used in \ac{MVDR}}
\Ac{MVDR} methods usually use contrastive loss as the training target, where negative documents are used.
For a query $q$ and target document $d$, the loss is computed as
\begin{equation}
    \mc{L}(d, q) =-\log\frac{\exp{\operatorname{rel}(d, q)}}{\sum_{d^-\in\mc{D}^-}\exp{\operatorname{rel}(d^-, q)}},
\end{equation}
where $\mc{D}^-$ is the collected negative set.

\section{In-Depth Analysis of Generative Retrieval}
\label{sec:brief_derivation}

To address the question posed in Section~\ref{sec:intro}, 
this section conducts a detailed analysis of GR.
Specifically, we first illustrate the model architecture and training loss of the GR (Section~\ref{sec:derivation_background}).
Subsequently, we derive that the training target of \ac{GR} falls into the framework of \ac{MVDR} (Section~\ref{sec:core_derivation}):
\begin{equation}
    \mc{L}(d, q)\propto \operatorname{sum}\left(\mat{\tilde{E}}_d^\top\mQ\odot\mA\right),
\end{equation}
where $\mat{\tilde{E}}_d$, $\mQ$ and $\mA$ correspond to $\mD$, $\mQ$ and $\mA$ in Eq.~\eqref{eq:framework_mvdr}.

\subsection{Model architecture and training loss}
\label{sec:derivation_background}
\header{Model architecture}
We focus on the transformer sequence-to-sequence architecture utilized in GR, more precisely, the encoder-decoder structure. 
Within this framework, the encoder primarily targets processing the input query, while the decoder is tasked with predicting document identifiers.

The decoder component consists of stacks of self-attention, cross-attention, and feed-forward layers. 
We particularly underscore the significance of the cross-attention layers, as they facilitate interaction between query tokens and document tokens.

To predict the document token $d_i$ at the $i$-th position, we compute the cross attention weights between query token vectors $\mQ$ and $\hat{\vd}_{i-1}$ from the previous attention layers at position $(i-1)$ as follows:
\begin{equation}
\vec{\alpha}_i = \operatorname{softmax}(\mQ^\top \mW \hat{\vd}_{i-1}),
\label{eq:gr_attn_weight_computing}
\end{equation}
where $\operatorname{softmax}(\cdot)$ denotes the column-wise softmax function, $\mW \in \R^{d \times d}$ is the product of the attention matrices $\mW_K$ and $\mW_Q$, i.e., $\mW = \mW_K^\top \mW_Q$, and $\vec{\alpha}_i \in \R^{N}$.

Consequently, the output of the cross-attention layer is
\begin{equation}
\vh_i = \mW_V\mQ\vec{\alpha}_i \in \R^{d}.
\label{eq:computing_h_i}
\end{equation}
For simplicity, we ignore the non-linear activation functions, and the linear maps in the feedforward layers can be absorbed in attention weights, e.g., $\mW_V$.
Therefore, $\vh_i$ serves as the prediction head for generating the next token. 

\header{Training loss}
The loss function to minimize at position $i$ is formulated as:
\begin{align}
\mc{L}_i(d, q) &= -\log\left(\frac{\exp{\ve_{d_i}^\top \vh_i}}{\sum_{v\in\V}\exp{\ve_v^\top \vh_i}}\right) \\
&= - \ve_{d_i}^\top \vh_i + \log \sum_{v\in\V}\exp{\ve_v^\top \vh_i}.
\end{align}

\subsection{\ac{GR} has the same framework as \ac{MVDR}}
\label{sec:core_derivation}

Next, we demonstrate that GR shares a similar framework with MVDR, namely, that the logits within the loss function can be reformulated as a product of document word embeddings, query token vectors, and attention matrix. This formulation corresponds to Eq.~\eqref{eq:framework_mvdr}.

In particular, as we employ teacher-forcing supervision, ground-truth document identifiers are directly fed into the decoder, and token vectors at all positions are computed simultaneously.
Based on this configuration, the overall loss is given by:
\begin{align}
    \mc{L}(d, q)
    & =\sum_{i\in[M]}\log p(d_i\mid d_{i-1},\ldots,d_0,q) \\
    & =\sum_{i\in[M]}\mc{L}_i(d, q) \\
    & =\sum_{i\in [M]} \left( - \ve_{d_i}^\top \vh_{i} + \log \sum_{v\in\V}\exp{\ve_v^\top \vh_i}\right),
    \label{eq:celoss}
\end{align}
where $d_0$ could be some special token such as \verb|[BOS]| or the \verb|[CLS]| token vector from the query.

When utilizing the sampled softmax loss, which involves employing several negative tokens instead of the entire set of tokens in the lookup embedding table, the loss exhibits a similar structure to the contrastive loss used in \ac{DR} and \ac{MVDR}. 
Consequently, we treat the dot product of embedding $\ve_{d_i}$ and token vector $\vh_i$, i.e., $\ve_{d_i}^\top \vh_i$, as the final relevance score at position $i$.
Further insights into the reason are elaborated in Appendix~\ref{app:sim_score_in_gr}.
We plug in the dot product with the computing of $\vh_i$ from Eq.~\eqref{eq:computing_h_i} and obtain:
\begin{align}
    \operatorname{rel}(d, q)&=\sum_{i\in[M]}\ve_{d_i}^\top \vh_i \\
    &=\sum_{i\in[M]}\ve_{d_i}^\top \mW_V\mQ\vec{\alpha}_i\\
    &=\sum_{i\in[M]}\sum_{j\in[N]}\vec{\tilde{e}}_{d_i}^\top \vq_j\alpha_{ij}\\
    &=\,\operatorname{sum}\left(\mat{\tilde{E}}_d^\top \mQ\odot\mA\right),
\end{align}
where $\vec{\tilde{e}}_{d_i}^\top=\ve_{d_i}^\top \mW_V$, 
$\mat{\tilde{E}}_d^\top=\mE_d^\top \mW_V$, 
$\mA=[\vec{\alpha}_1,\ldots,\vec{\alpha}_{M}]^\top\in\R^{M\times N}$,
and $\odot$ is the element-wise matrix product operation.

Further, we have a more detailed computation
\begin{align}
    \operatorname{rel}(d, q)&=\operatorname{sum}\left(\mat{\tilde{E}}_d^\top \mQ\odot\mA\right)\\
    &=\operatorname{sum}\left(\mat{\tilde{E}}_d^\top \mQ\odot\operatorname{softmax}\left( \mat{\hat{D}}_{-1}^\top\mW^\top\mQ\right)\right),
    \label{eq:gr_compact_framework_target}
\end{align}
where $\mat{\hat{D}}_{-1}=\left[\vec{\hat{d}}_0,\vec{\hat{d}}_1,\ldots,\vec{\hat{d}}_{M-1}\right]$ is the output from the previous layer with the right-shifted document tokens as model input.

\begin{figure}[t]
    \centering
    \includegraphics[width=\linewidth]{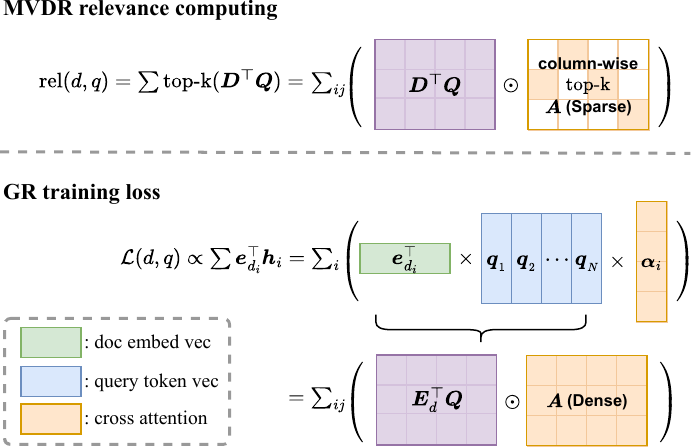}
    \caption{Summary of our derivation and conclusion.
    The logits of \ac{GR} can be reformulated as $\operatorname{sum}(\mE_d^\top\mQ\odot\mA)$, which corresponds to the framework $\operatorname{sum}(\mD^\top \mQ \odot \mA)$ of \ac{MVDR}.}
    \label{fig:main_conclusion}
\end{figure}

In conclusion, we observe a similar framework from \ac{GR} to \ac{MVDR}, $\operatorname{rel}(d, q)=\operatorname{sum}(\mat{\tilde{E}}_d^\top \mQ\odot \mA)$, where the relevance is represented by an interaction of multiple ``token vectors,'' i.e., $\mat{\tilde{E}}_d$ and $\mQ$, from both query and document and aligned by a matrix $\mA$.
We summarize our derivation and conclusion in Figure~\ref{fig:main_conclusion}.

\section{Comparison between \ac{MVDR} and \ac{GR}}
\label{sec:main_comparison}

To further explore how GR is related to MVDR, we build upon the unified framework of relevance computation for GR and MVDR derived in the previous section.
We conduct a comprehensive analysis of both methods, focusing specifically on their similarities and disparities in terms of the document encoding and the design of the alignment matrix.
A summary of the comparison between the two methods is shown in Table~\ref{table:formula_comparison}.

\subsection{Document encoding}
\label{sec:compari_document_encoding}

One of the noticeable differences between GR and MVDR is in the document encoding.
As depicted in Figure~\ref{fig:main_conclusion}, \ac{MVDR} utilizes more expressive contextualized token vectors 
$\mD=\left[ \vd_1, \ldots, \vd_M \right]$ for each position. In contrast, \ac{GR} only attends each query token to a simple word embedding $\ve_{d_i}$ that does not hold any contextual information about the document. This was considered a severe compromise for the extremely lightweight modeling and storage of \ac{GR}.
To address this imbalance in modeling capacity, several studies \citep{leeNonparametricDecodingGenerative2023,wangNeuralCorpusIndexer2023} have proposed novel decoding methods.
\citet{wangNeuralCorpusIndexer2023} introduce the \acf{PAWA} decoding method, while \citet{leeNonparametricDecodingGenerative2023} propose the \acf{NP} decoding.
We incorporate these methods into our framework and show how they fundamentally improve the encoding compared with \ac{MVDR} in Table~\ref{table:document_encoding_comparison}.

\begin{table}[t]
\caption{Summary of our comparison  between \ac{MVDR} and \ac{GR}.}
\label{table:formula_comparison}
\centering
\setlength{\tabcolsep}{1.1mm}
\begin{tabular}{p{2.2cm}lll}
    \toprule
    \textbf{Component in} & \multicolumn{2}{c}{\textbf{Model}} & \multirow{2}*{\begin{tabular}{@{}l@{}}\textbf{Comp-} \\\textbf{arison}\end{tabular}}\\
    \cmidrule(lr){2-3}
      $\mathrm{sum}(\textcolor{blue}{\mat{D}}^\top \mat{Q}\odot \textcolor{red}{\mat{A}})$ & \ac{MVDR} (Sect.~\ref{sec:MVDR}) & \ac{GR} (Sect.~\ref{sec:brief_derivation}) & \\
    \midrule
    \textcolor{blue}{$\mat{D}$ doc token}
    & $\mat{D}$ (token vec.) & $\mat{\tilde{E}}_d$ (embed. vec.) & Sect.~\ref{sec:compari_document_encoding}\\
    
    $\mat{Q}$ query token
    & $\mat{Q}$ (token vec.) & $\mat{Q}$ (token vec.) & --- \\
    
    \addlinespace
    \multirow{2}*{\begin{tabular}{@{}l}\textcolor{red}{$\mat{A}$ alignment}\\ \textcolor{red}{matrix}\end{tabular}}
    & sparse & dense and learned & \multirow{2}*{Sect.~\ref{sec:compari_alignment_strategy}} \\
    & query-to-doc & doc-to-query & \\
    \bottomrule
\end{tabular}
\end{table}

\begin{table}[ht]
\caption{Document encoding comparison between \ac{GR} and \ac{MVDR}.
\ac{PAWA} and \ac{NP}-decoding either multiply or replace the simple embedding vectors 
$\mat{\tilde{E}}_d$ with contextualized token vectors $\mat{\tilde{D}}$.}
\label{table:document_encoding_comparison}
\centering
\begin{tabular}{ll}
    \toprule
    \textbf{Model} & \textbf{Document encoding} \\
    \midrule
    \ac{MVDR} (Sect.~\ref{sec:MVDR}) & $\mat{D}$ (token vec.) \\
    \ac{GR} (Sect.~\ref{sec:brief_derivation}) & $\mat{\tilde{E}}_d$ (embed. vec.) \\
    -- w/ \ac{PAWA} & $\mat{\tilde{E}}_d\rightarrow\mat{\tilde{E}}_d\mat{\tilde{D}}^{\prime}$ (embed. \& token vec.) \\
    -- w/ \ac{NP}-dec. & $\mat{\tilde{E}}_d\rightarrow\mat{D}$ (token vec.)\\
    \bottomrule
\end{tabular}
\end{table}

\header{\ac{PAWA} enhances the document encoding from $\mat{\tilde{E}}_d$ to $\mat{\tilde{E}}_d\mat{\tilde{D}}'$}
\label{sec:PAWA}
\ac{PAWA} \citep{wangNeuralCorpusIndexer2023} aims to improve the embedding modeling for distinguishing different semantics of a token ID at different positions.
Unlike the standard transformer, which uses a static embedding lookup table for every position, \ac{PAWA} generates different embedding tables at each generation step.
\ac{PAWA} consists of a transformer decoder and an adaptive projection layer $\tensor{E}\in\R^{M\times\nV\times d\times d}$.
The projection matrix of token $v$ at the $i$-th position is
\begin{equation*}
\mE_{i,v}=\tensor{E}[i,v,:,:]\in\R^{d\times d}.
\end{equation*}
Here, $\tensor{E}$ can be seen as a generalized version of the embedding lookup table that uses a matrix $\mE_{i,v}$ to represent each token $v$.
To get the generated embedding vector for token $v$ at the $i$-th position, \ac{PAWA} decoder first uses the transformer decoder to process the document into a set of latent vectors $\mD'=[\vd'_1,\ldots,\vd'_M]\in\R^{d\times M}$.
Then it multiplies the projection matrix $\mE_{i,v}$ with the latent vector $\vd'_i$ and gets the final embedding vector $\ve_{i,v}=\mE_{i,v}\vd'_i$.

Therefore, we have the logit $\ve_v^\top\vh_i$ in loss Eq.~\eqref{eq:celoss} replaced by $\ve_{i,v}^\top\vh_i=\vd_i^{\prime\top}\mE_{i,v}^\top\vh_i$:
\begin{equation}
    \mc{L}(d, q)=\sum_{i\in [M]} \left[ - \vd_i^{\prime\top}\mE_{i,d_i}^\top \vh_{i} + \log \sum_{v\in\V}\exp{\vd_i^{\prime\top}\mE_{i,v}^\top \vh_i}\right].
    \label{eq:pawa_ce_loss}
\end{equation}
With a similar derivation as in Section~\ref{sec:core_derivation}, the relevance can be established as
\begin{equation}
    \operatorname{rel}(d, q)=\operatorname{sum}\left(\mat{\tilde{D}}^{\prime\top}\mat{\tilde{E}}_d^\top \mQ\odot\mA\right),
\end{equation}
where 
\begin{equation}
    \mat{\tilde{D}}^{\prime} = 
    \begin{bmatrix}
    \vd'_1 & \mathbf{0} & \cdots & \mathbf{0} \\
    \mathbf{0} & \vd'_2 & \cdots & \mathbf{0} \\
    \vdots & \vdots & \ddots & \vdots \\
    \mathbf{0} & \mathbf{0} & \cdots & \vd'_M
    \end{bmatrix}
    \in\R^{(d\times M)\times M},
\end{equation} 
and
$\mat{\tilde{E}}_d=\mW_V^\top[\mE_{1,d_1},\ldots,\mE_{M,d_M}]\in\R^{d\times (d\times M)}$.

As we can see, \ac{PAWA} multiplies the term $\mat{\tilde{E}}_d^\top \mQ$ with contextualized document token vectors $\mat{\tilde{D}}^{\prime}$, which greatly improves the expressivity of document encoding.

\header{\Ac{NP}-decoding directly replaces $\mat{\tilde{E}}_d$ with $\mat{D}$}
\label{sec:np_decoding}
\citet{leeNonparametricDecodingGenerative2023} employ an approach akin to contextualized sparse retrieval methods, leveraging token vectors encoded by the Contextualized Embedding Encoder \citep[CE Encoder,][]{leeNonparametricDecodingGenerative2023}, referred to as contextualized token embeddings.
This set of vectors, denoted as $\mD$ in our notation, serves as the embedding table for the decoder.
Both the \textsc{Base} and \textsc{Async} nonparametric decoding methods in~\citep{leeNonparametricDecodingGenerative2023} can be reformulated within our framework as:
\begin{equation}
    \operatorname{rel}(d, q)=\operatorname{sum}(\mat{D}^\top \mQ\odot \mA),
\end{equation}
where $\mat{D}$ is the token vectors of documents either pre-computed (as done by the pre-trained T5 model and frozen in the \textsc{Base} method) or gradually updated (by the encoder of the \ac{GR} model every $N$ epochs in the \textsc{Async} method) during the training of the \ac{GR} model.

While the \ac{NP}-decoding method shares the same document encoding with \ac{MVDR}, two significant differences exist:

\begin{enumerate}[leftmargin=*]
    \item In \ac{NP}-decoding, $\mat{D}$ is mostly frozen and detached from training, causing training imbalance compared to \ac{MVDR}. 
    $\mQ$ and $\mA$ in \ac{NP}-decoding are fully trained, while $\mD$ remains frozen.
    \item Due to \ac{GR} computing logits for the entire vocabulary in each generation step, there's a need to reduce the large storage footprint of $\mD$ to save computation. 
    \ac{NP}-decoding methods address this by using clustering to compress token vectors. 
    \ac{MVDR}, on the other side, achieves lower inference time through a sparse alignment strategy.
\end{enumerate}

\subsection{Alignment strategy}
\label{sec:compari_alignment_strategy}

In addition to document encoding, the alignment matrix represents a crucial distinction between \ac{GR} and \ac{MVDR} methods.
This matrix plays a decisive role in shaping the divergent inference procedures employed in retrieval.
In this section, we analyze the alignment matrix, denoted as $\mA$ within our unified framework, in terms of sparsity, alignment direction, and some common properties.

\subsubsection{The concept of ``alignment'' in both methods}
\label{sec:alignment_concept_in_both}
The concept of ``alignment'' has garnered significant attention in \ac{MVDR}~\citep{qianMultiVectorRetrievalSparse2022,leeRethinkingRoleToken2023,liCITADELConditionalToken2022,coilcr2023}. 
We will briefly introduce the alignment problem in \ac{MVDR} models, and claim that the alignment matrix of the \ac{GR} method, as asserted in our framework Eq.~\eqref{eq:gr_compact_framework_target}, indeed exhibits similar alignment functionality to \ac{MVDR} models.

Token alignment involves determining whether tokens from the query and document should be matched lexically or semantically.
It essentially represents another formulation of the ``term mismatch problem''~\citep{qianMultiVectorRetrievalSparse2022,zhaoTermMismatch2012,khattabColBERTEfficientEffective2020,formalSPLADESparseLexical2021}.
The prevailing strategy considered optimal at present is the all-to-all soft alignment strategy in \ac{MVDR} models~\citep{khattabColBERTEfficientEffective2020,qianMultiVectorRetrievalSparse2022}, which eliminates the lexical form match restriction.

\ac{GR} methods leverage the transformer architecture that originated in NLP, and the concept of ``alignment'' has been extensively discussed in the domain of neural machine translation~\citep{chenAccurateWordAlignment2020,liWordAlignmentEra2022,liStructuralSupervisionWord2022,chen2021lexically}, focusing on the alignment between tokens in source and target sentences.
The attention mechanism, as a core component, computes the alignment matrix and proves highly effective in capturing alignment between source and target sentences~\citep{chenAccurateWordAlignment2020,Zhang2021MindTG,jiang-etal-2020-cross, Yu2021CrosslingualLM}.
Theoretical works~\citep{elhage2021mathematical,olsson2022context} have further validated the phenomenon of copying behavior, forming a foundational basis for the alignment ability.

We conclude that, in \ac{GR} methods, the attention matrix is able to capture the alignment between the query and the document, akin to the alignment matrix observed in \ac{MVDR} methods.

\subsubsection{Different sparsity and learnability: sparse vs. dense and learned alignment matrices}
The alignment matrices of \ac{MVDR} and \ac{GR} differ in sparsity and learnability. 
\ac{MVDR} typically employs a sparse alignment matrix for maximum efficiency during inference.
In contrast, \ac{GR} utilizes a dense and fully learnable alignment matrix derived from the computationally intensive attention mechanism.

\textbf{For \ac{MVDR} methods}, the sparse alignment matrix is often computed using heuristic algorithms~\citep{khattabColBERTEfficientEffective2020,gaoCOILRevisitExact2021,qianMultiVectorRetrievalSparse2022}.
Taking ColBERT~\citep{khattabColBERTEfficientEffective2020} as an example, it selects the most relevant document token for each query token.
The relevance score between document $d$ and query $q$ is computed as
\begin{align}
    \operatorname{rel}(d, q) &= 
    \operatorname{sum}\left(\mD^\top\mQ\odot\mA\right) = \sum_{i,j}\vd_i^\top\vq A_{ij}\nonumber\\
    &=\sum_{j\in[N]}\max_{i\in[M]}\vd_i^\top \vq_j = \sum_{j\in[N]}\operatorname{rel}(d, q_j),
    \label{eq:mvdr_sim_decomposition}
\end{align}
where $\mA$ is a sparse alignment matrix with only one non-zero element for each column ($A_{ij}=1$ if $\vd_i^\top \vq_j=\max_{i\in[M]}\vd_i^\top \vq_j$; $A_{ij}=0$ otherwise).
The sum-max operation is highly parallelizable, ensuring efficiency during inference.

\textbf{For \ac{GR} methods}, the alignment matrix is computed through the attention mechanism, considering all possible pairs of query and document tokens, as shown in Eq.~\eqref{eq:gr_compact_framework_target}.

The dense alignment matrix is highly expressive and trainable.
While not suitable for exact relevance score computation in inference for each query-document pair, efficient approximate algorithms such as greedy search or beam search can be used to retrieve the top-$k$ documents.
These decoding algorithms rely on the following decomposition:
\begin{align}
    \operatorname{rel}(d, q) 
    = \sum_{i\in[M]}\operatorname{sum}\left(\vec{\tilde{e}}_{d_i}^\top\mQ\vec{\alpha}_i\right) = \sum_{i\in[M]}\operatorname{rel}(d_i, q),
    \label{eq:gr_sim_decomposition}
\end{align}
where $\operatorname{rel}(d_i, q)$ is conditioned on $d_0,\ldots,d_{i-1}$, approximating the search for the most relevant document $d$ to finding the most relevant token $d_i$ at each position $i$.

\subsubsection{Different alignment directions: query-to-document vs.\ docu\-ment-to-query alignment}
\label{sec:alignment_direction}

Beyond differences in the sparsity and learnability of the alignment matrix, \ac{MVDR} and \ac{GR} exhibit distinctions in their alignment directions.

Eq.~\eqref{eq:mvdr_sim_decomposition} reveals that MVDR's relevance score can be decomposed into the sum of relevance scores for each query token and its aligned document token.
In this context, we consider the alignment matrix in \ac{MVDR} as \textbf{query-to-document} alignment.
Each query token individually aligns to a document token, seeking the optimal match during retrieval.
Mathematically, the alignment matrix is computed \emph{column-wise} and represents a one-hot vector for each column.

Conversely, the relevance score of \ac{GR}, as depicted in Eq.~\eqref{eq:gr_sim_decomposition}, is the sum of relevance scores for each document token and its softly aligned query token.
Here, we categorize the alignment matrix in \ac{GR} as \textbf{document-to-query} alignment.
Each document token is considered individually to focus attention on the most relevant query token.
The alignment matrix is computed \emph{row-wise} with a $\operatorname{softmax}(\cdot)$ operation to normalize attention weights in each row.

Document-to-query alignment may seem counter-intuitive for a retrieval task, as we do not know the target documents while predicting.
As a solution, \ac{GR} pre-computes the alignment strategy for the document token $d_i$ (to be predicted) using \emph{previous} document tokens $d_0,\ldots,d_{i-1}$ and thus can retrieve the next token that best aligns with the desired \emph{next} alignment strategy.

\subsubsection{Low-rank nature of both alignment matrices}
\label{sec:low_rank_property}
In analyzing the shared characteristics of the two alignment matrices, it is demonstrated that both matrices exhibit a low-rank property.

\ac{MVDR} model, e.g., \textsc{AligneR}~\citep{qianMultiVectorRetrievalSparse2022}, integrates the pairwise alignment matrix with unary salience, given by
\begin{equation}
    \mA=\mat{\tilde{A}}\odot\vec{u}_d\vec{u}_q^\top.
    \label{eq:aligner_alignment_matrix}
\end{equation}
Here, $\mat{\tilde{A}}\in\R^{M\times N}$ signifies the pairwise alignment matrix, determining the alignment of query and document tokens.
The sparse token weights, $\vec{u}_d\in\R^M$ and $\vec{u}_q\in\R^N$, decide whether a token requires alignment.
Notably, the alignment matrix $\mA$ contains a low-rank component $\vec{u}_d\vec{u}_q^\top$ that influences the alignment strategy.

In the case of \ac{GR} methods, the alignment matrix is computed using an attention mechanism, which inherently results in a low-rank matrix.
A lemma provides evidence of this low-rank property and is presented briefly here, with a detailed proof in Appendix~\ref{app:gr_alignment_low_rank}.
\begin{lemma}
    For a matrix $\mA = \operatorname{softmax}\left(\mD^\top \mW\mQ\right)$, there exists a rank-one matrix $\mat{R}$ such that
    \begin{equation}
        \|\mA-\mat{R}\|\le 4\gamma \|\mW\|,
    \end{equation}
    where the term $\gamma$ depends on the matrix entries.
    \label{lemma:gr_alignment_low_rank}
\end{lemma}

\noindent%
From this lemma, we can conclude that both \ac{MVDR} and \ac{GR} methods reveal a rank-one component in their alignment matrices.

\subsubsection{Decomposition of both relevance scores}
In this subsection, we show that the relevance score computation in both \ac{MVDR} and \ac{GR} models can be decomposed into query and document components.

The \ac{MVDR} method employs a bi-encoder architecture, wherein query and document tokens are modeled separately.
This architecture can easily be regarded as a decomposition of the relevance score between the query and document:
\begin{align*}
    \operatorname{rel}(d, q) = \operatorname{sum} \left(\mD^\top\mQ\odot\mA\right)
    = \operatorname{sum} \left(\operatorname{top-1}( \mD^\top \mQ)\right),
\end{align*}
where $\operatorname{top-1}(\cdot)$ is the operator that selects the maximum value in each column of the matrix.

In the subsequent lemma, we establish that the relevance score of \ac{GR} cannot only be decomposed but also be kernelized, implying the existence of a kernel function capable of processing both query vectors and document vectors to compute the score (further details can be found in Appendix~\ref{app:gr_alignment_kernel}):
\begin{lemma}
    For simplicity, let $\operatorname{rel}(d, q) = \operatorname{sum}(\mD^\top \mQ\odot \mA)$, where $\mA= \operatorname{softmax}(\mD^\top\mQ)$.
    It can be kernelized as
    \begin{align}
        \operatorname{rel}(d, q) &= \sum_{i,j} \vd_i^\top \vq_j A_{ij}=\sum_{i,j} \vd_i^\top \vq_j \operatorname{softmax}\left(\vd_i^\top \mQ\right)_j \nonumber\\
        &= \sum_{i,j}\frac{1}{p_{ij}}\operatorname{tr}\left(\mat{F}(\vd_i)^\top \mat{F}(\vq_j)\right),
    \end{align}
    where $\mat{F}(\vec{x})=\vec{x}\vec{\phi}(\vec{x})^\top$, and $p_{ij}$ is a term that depends on $\vd_i$ and $\vq_j$.
    We choose $\operatorname{elu}(\cdot)$ as the kernel function $\vec{\phi}(\cdot)$.

    Furthermore, by applying the trace inequality, we can approximately decompose the relevance score as
    \begin{equation*}
        \operatorname{rel}(d, q) \le \sum_{i,j}\frac{1}{\hat{p}_{i}\hat{p}_{j}}\sqrt{\operatorname{tr}\left(\mat{F}(\vd_i)^\top\mat{F}(\vd_i)\right)}\sqrt{\operatorname{tr}\left(\mat{F}(\vq_j)^\top\mat{F}(\vq_j)\right)}.
    \end{equation*}
    \label{lemma:gr_alignment_kernel}
\end{lemma}

\noindent%
From this lemma, we can conclude that both relevance scores in \ac{MVDR} and \ac{GR} methods can be decomposed.
The decomposition of \ac{MVDR} is more straightforward, and the kernelization of \ac{GR} is more complicated.
Both kernelizations would provide possibilities for new retrieval strategies.

\subsection{Upshot}
In summary, our findings indicate that certain studies enhance the modeling capacity of \ac{GR} by employing more expressive document encoding, akin to \ac{MVDR}. 
Furthermore, \ac{GR} employs a distinct alignment direction, but it also exhibits similar low-rank and decomposition properties with \ac{MVDR}.

\section{Experimental setup}

Next, we seek experimental confirmation that generative retrieval and multi-vector dense retrieval share the same framework for measuring relevance to a query of a document, as derived in Section~\ref{sec:brief_derivation}.

\subsection{Datasets}
We conduct experiments on two well-known datasets, NQ~\citep{nq-dataset} and MS MARCO~\citep{msmarco-dataset}.
We use the same settings and processed datasets as~\citet{sunLearningTokenizeGenerative2023}, and we summarize the statistics of the datasets in Table~\ref{table:data_stats}.

\begin{table}[ht]
\centering
\setlength\tabcolsep{5pt}
\caption{Statistics of datasets used in our experiments.}
\label{table:data_stats}
\begin{tabular}{l ccc}
\toprule
Dataset & \# Docs & \# Test queries & \# Train pairs\\
\midrule
NQ320K & 109,739 & 7,830  & 307,373 \\
MS MARCO & 323,569 & 5,187 & 366,235 \\
\bottomrule
\end{tabular}
\end{table}

\header{NQ320k}
NQ320K is a popular dataset for evaluating retrieval models~\citep{karpukhinDensePassageRetrieval2020,tayTransformerMemoryDifferentiable2022a,wangNeuralCorpusIndexer2023,liLearningRankGenerative2023}.
It is based on the Natural Questions (NQ) dataset~\citep{nq-dataset}.
NQ320k consists of 320k query-document pairs, where the documents are gathered from Wikipedia pages, and the queries are natural language questions.

\header{MS MARCO}
MS MARCO document retrieval dataset is a collection of queries and web pages from Bing searches.
Like NQ320k and following~\citep{sunLearningTokenizeGenerative2023}, we sample a subset of documents from the labeled documents and use their corresponding queries for training.
We evaluate the models on the queries of the MS MARCO dev set and retrieval on the sampled document subset.

\subsection{Base models}
As we aim to provide a new perspective on \ac{GR} as \ac{MVDR}, we consider representative models from both paradigms, i.e., SEAL~\citep{seal2022} for \ac{GR} and ColBERT~\citep{khattabColBERTEfficientEffective2020} for \ac{MVDR}.
For a fair comparison, we reproduce both methods using the T5 architecture~\citep{t52020}.
We have made several changes to adapt ColBERT and SEAL to their T5 variants:\footnote{Our code link is \url{https://github.com/Furyton/GR-as-MVDR}.}

\begin{itemize}[leftmargin=*,nosep]
\item \textbf{T5-ColBERT.}
We use in-batch negative samples instead of the pair-wise samples in the official ColBERTv1 implementation.
We set the batch size to 256 and train 5 epochs.
The details of our T5 variant ColBERT are in Appendix~\ref{app:detail_of_T5_variants}.
\item \textbf{T5-SEAL.}
We use the Huggingface transformers library~\citep{huggingface} to train the model.
We use the constructed query-to-span data for training and each span has a length of 10 sampled according to \citet{seal2022}.
The learning rate is set to 1e-3 and the batch size is 256.
\end{itemize}

\subsection{Inference settings}
We consider two inference settings: end-to-end and reranking.

\header{End-to-end retrieval setting}
Both methods can perform an end-to-end retrieval on the corpus for a given query.
\begin{itemize}[leftmargin=*,nosep]
    \item \textbf{T5-ColBERT} maintains a large vector pool of all document token vectors after training.
    During inference, it first retrieves for each query token vector, the $k$-nearest document token vectors in the vector pool, resulting in $N\times k$ retrieved vectors.
These vectors are from at most $N\times k$ different documents which are used as candidates.
It then computes the exact relevance score for each candidate document and performs the final rerank.
    \item \textbf{T5-SEAL} directly uses its generative style inference with the help of constrained beam search to predict valid document identifiers, i.e., n-grams from the documents.
\end{itemize}

\header{Rerank setting}
Since we are focusing on relevance computing in the training target, we introduce a rerank setting that removes the influence of different approximated inference strategies.
As stated in some previous work~\citep{leeRethinkingRoleToken2023,liLearningRankGenerative2023}, both \ac{MVDR} and \ac{GR} have discrepancies between training and inference.
The approximated retrieval methods are largely different from the training target and may decrease the performance of the trained retrievers.
In the rerank setting, we collect 100 documents retrieved by BM25~\citep{bm25} together with the ground-truth document as the candidate set for each query.
As in the training stage, we take both the query and each candidate document as the input of the model and use the relevance computing in~Section~\ref{sec:MVDR} and~\ref{sec:brief_derivation}.

\section{Experimental analyses}
\subsection{Performance of different alignment directions}
As described in Section~\ref{sec:alignment_direction}, \ac{MVDR} and \ac{GR} exhibit different alignment directions, i.e., query-to-document and document-to-query alignment.
We aim to look at how alignment directions affect retrieval performance.
We first conduct experiments in the rerank setting to show the performance gap between \ac{MVDR} and \ac{GR}.
As shown in Table~\ref{table:rerank_performance}, \ac{MVDR} with the original alignment strategy, which is indicated as \ac{MVDR}~(q$\rightarrow$d), has a much better performance than \ac{GR}.
To compare the alignment directions of \ac{MVDR} and \ac{GR}, we have designed a model \ac{MVDR}~(q$\leftarrow$d) that integrates the features of both, i.e., expressive document encoding from \ac{MVDR} and document-to-query alignment strategy from \ac{GR}.
From Table~\ref{table:rerank_performance}, we can see that the performance of the new model is roughly intermediate between the other two.
Note that the designed experimental model \ac{MVDR}~(q$\leftarrow$d) can only be used in a rerank setting.
We conclude that query-to-document alignment is preferred for reranking.

\begin{table}[t]
\caption{Comparison of \ac{MVDR} and \ac{GR} in rerank setting.
\ac{MVDR} and \ac{GR} are our reproduced T5-SEAL and T5-ColBERT.
``R'' denotes Recall, and ``M'' denotes MRR.
}
\label{table:rerank_performance}
\begin{tabular}{@{}l cccccc@{}}
    \toprule
     & \multicolumn{3}{c}{\textbf{NQ320K}} & \multicolumn{3}{c}{\textbf{MS MARCO}} \\
    \cmidrule(lr){2-4}    \cmidrule(lr){5-7}
    \textbf{Model} & R@1 & R@10 & M@10 & R@1 & R@10 & M@10 \\
    \midrule
    \ac{MVDR}~(q$\rightarrow$d)  & 61.3 & 91.9 & 72.0    & 46.5 & 84.5 & 58.9 \\
    \ac{MVDR}~(q$\leftarrow$d) & 53.2 & 90.1 & 65.7 & 34.8 & 78.8 & 48.4 \\
    \ac{GR} & 47.4 & 87.0 & 60.5    & 35.3 & 77.1 & 48.3 \\
    \bottomrule
\end{tabular}
\end{table}

\subsection{Term matching in alignment}
As we have discussed in Section~\ref{sec:alignment_concept_in_both}, alignment is essentially a term-matching problem.
In this section, we design an experiment to observe the extent of term matching in the two methods, and we find both methods exhibit a preference for exact term matching in their alignment.

\begin{figure}[h]
    \centering
    \includegraphics[width=\linewidth]{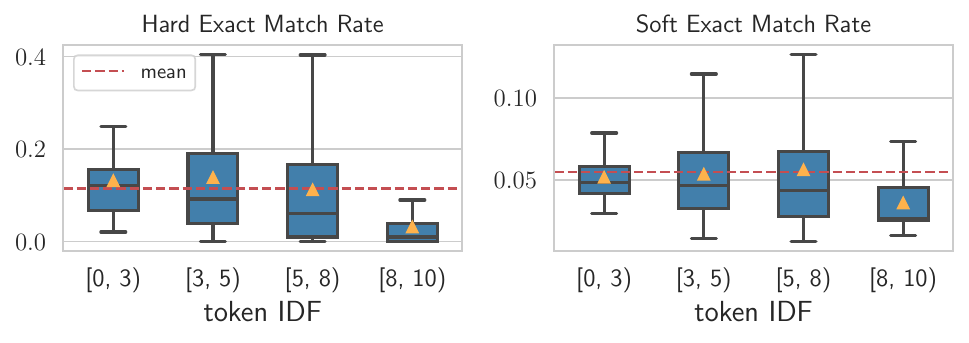}
    \caption{Exact match rate of \ac{MVDR} on NQ320k dataset in query-to-document direction.}
    \label{fig:hsemr_nq_mvdr_q2d}
\end{figure}

\header{Exact match of \ac{MVDR} in query-to-document direction}
We calculate the exact matching rate between document token IDs and each query token ID during the alignment process, which we refer to as the ``hard exact match rate.''
We also define a ``soft exact match rate'' which is the alignment score corresponding to the exact match query-document token pairs.
The alignment score is defined as the element in the alignment matrix.
As \ac{MVDR} uses a sparse alignment matrix, we apply column-wise $\operatorname{softmax(\cdot)}$ to $\mA$ and use the element as the alignment score.
We average the rate over candidate documents for each query token and categorize the query tokens according to their IDF.
We assume that IDF approximates the term importance as is done in~\citep{analysis-of-colbert}.
From the result in Figure~\ref{fig:hsemr_nq_mvdr_q2d}, we can see that \ac{MVDR} chooses exactly matched document tokens in 11.4\% on average.
Also, we notice that rare query tokens have not received much attention during alignment.
This observation suggests that \ac{MVDR} may prioritize commonly occurring query tokens in its alignment process, potentially overlooking or underemphasizing the importance of rare query tokens.

\begin{figure}[ht]
    \centering
    \includegraphics[width=\linewidth]{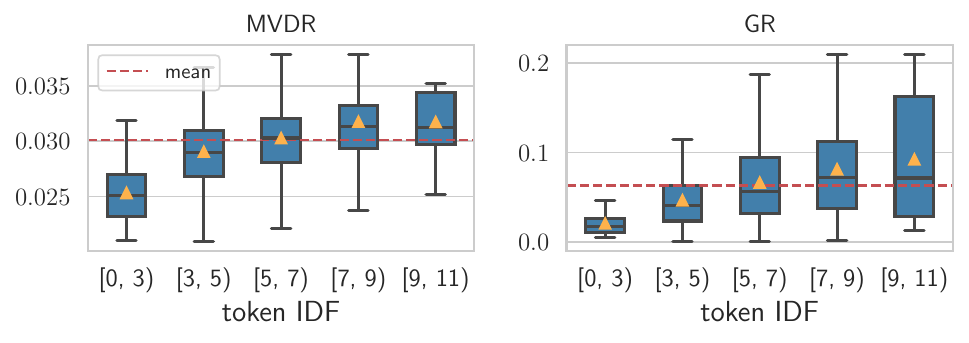}
    \caption{Soft exact match rate of \ac{MVDR} and \ac{GR} on NQ320k dataset in document-to-query direction.}
    \label{fig:semr_nq_mvdr_gr_d2q}
\end{figure}

\header{Exact match of \ac{MVDR} and \ac{GR} in document-to-query direction}
We have devised an experiment to investigate the alignment in the opposite direction, i.e., document-to-query, in Figure~\ref{fig:semr_nq_mvdr_gr_d2q}.
As \ac{GR} is not trained with hard alignment, we only examine the soft exact match rate of both methods.
The computation of the exact match rate is similar except that it is computed for each document token.
From the results, we have discerned a consistent trend: as the importance of tokens increases, the rate of exact matches also tends to rise.
We think this is because it is hard for the rare query token to match among many common tokens since the document is much longer than the query.
When we look at each document token, it will be easier to match among fewer query tokens.
We also conduct experiments on MS MARCO and have similar results.

\subsection{Improved document encoding}
In Section~\ref{sec:compari_document_encoding}, we include two popular document encoding methods, \ac{PAWA} and \ac{NP}-decoding, into our framework.
To demonstrate the improvement of these two methods, we compare the performance of \ac{GR} with and without them in Table~\ref{table:document_encoding_performance}.
\Ac{PAWA} is typically used in \ac{GR} with short semantic identifiers due to its high computational complexity during generation.
Thus, we compare it with DSI~\citep{tayTransformerMemoryDifferentiable2022a} with semantic identifiers.
Note that all these models use the same architecture (T5-base~\citep{t52020}) and similar training procedures without data augmentation, e.g., synthetic query-doc pairs generation, etc.
*-\ac{PAWA} and *-\ac{NP} can be seen as a naive approach to using the two enhancing methods to the base \ac{GR} models.
From Table~\ref{table:document_encoding_performance}, we see that both \ac{PAWA} and \ac{NP}-decoding can greatly improve the performance and achieve similar results compared with T5-ColBERT on Recall@1.
However, there is still a large gap in terms of Recall@10.
The implementation of the additional decoding modules is only an approximation for reducing the cost of time and storage as discussed in Section~\ref{sec:compari_document_encoding}.
This, together with the alignment direction, may be a cause of the performance gap between \ac{GR} equipped with these document encodings and \ac{MVDR}.

\begin{table}[ht]
\caption{Performance of \ac{GR} models with different document encoding methods on NQ320k in end-to-end setting.
The results of DSI-PAWA, DSI, and T5-GENRE-\ac{NP} are from~\citep{wangNeuralCorpusIndexer2023,leeNonparametricDecodingGenerative2023,sunLearningTokenizeGenerative2023}.
T5-GENRE is the T5 variant of GENRE~\citep{decaoAutoregressiveEntityRetrieval2020} used by~\citep{leeNonparametricDecodingGenerative2023}.}
\label{table:document_encoding_performance}
\begin{tabular}{l ccc}
    \toprule
    \textbf{Model} & R@1 & R@10 \\
    \midrule
    T5-ColBERT  & 61.1 & 88.4 \\
    \midrule
    DSI~\citep{tayTransformerMemoryDifferentiable2022a} & 55.2 & 67.4 \\
    DSI-PAWA~\citep{wangNeuralCorpusIndexer2023} & 60.2 & 80.2 \\
    \midrule
    T5-SEAL & 44.7 & 75.5 \\
    T5-GENRE~\citep{decaoAutoregressiveEntityRetrieval2020} & 53.7 & 64.7 \\
    T5-GENRE-\ac{NP}~\citep{leeNonparametricDecodingGenerative2023} & 62.2 & 78.8 \\
    \bottomrule
\end{tabular}
\end{table}

\subsection{Low-rank nature of alignment matrix}
\begin{figure}[t]
  \centering
  \subfigure[Relation of $\|\mat{W}\|$ and $\|\mat{A}-\mat{R}\|$]{\includegraphics[width=0.485\columnwidth]{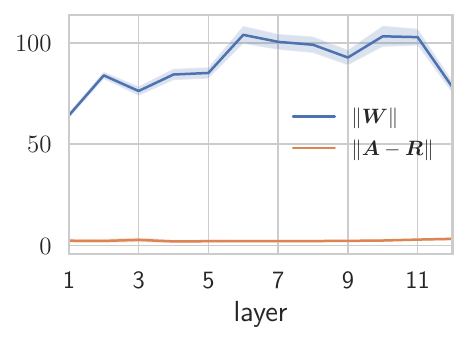}
  \label{fig:rank_analysis-norm}}
  \hfill
  \subfigure[Relative error rate of $\mat{R}$]{\includegraphics[width=0.485\columnwidth]{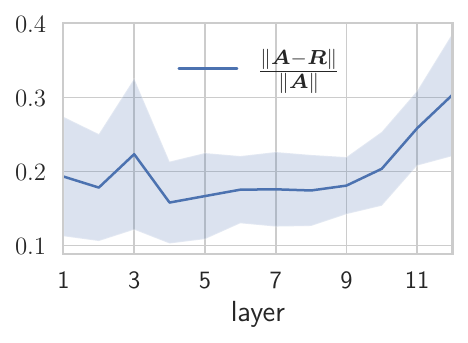}
  \label{fig:rank_analysis_rate}}
  \caption{Low-rank approximation of $\mat{R}$ to alignment matrix $\mat{A}$ in \ac{GR} in MS MARCO dataset.}
  \label{fig:rank_analysis}
\end{figure}

In Section~\ref{sec:low_rank_property}, we show that the alignment matrix in \ac{GR} also has a low-rank property in Lemma~\ref{lemma:gr_alignment_low_rank}.
As \ac{MVDR} using alignment matrix~\eqref{eq:aligner_alignment_matrix} already contains a low-rank component, we only conduct experiments to verify \ac{GR}.
Since the $\gamma$ in Lemma~\ref{lemma:gr_alignment_low_rank} is hard to attain, we illustrate the relation between $\|\mW\|$ and $\|\mA-\mat{R}\|$ in Figure~\ref{fig:rank_analysis-norm}.
We can see that the inequality is loose and $\|\mA-\mat{R}\|$ is much lower than $\|\mW\|$.
We also show the relative error of the approximation of $\mat{R}$ in Figure~\ref{fig:rank_analysis_rate}.
The error is relatively low on average, which indicates the low-rank nature of the alignment matrix of \ac{GR}.

\subsection{Case study of the alignment matrix}
\begin{figure}[t]
  \centering
  \subfigure[\ac{MVDR} document-to-query direction]{\includegraphics[height=0.43\linewidth]{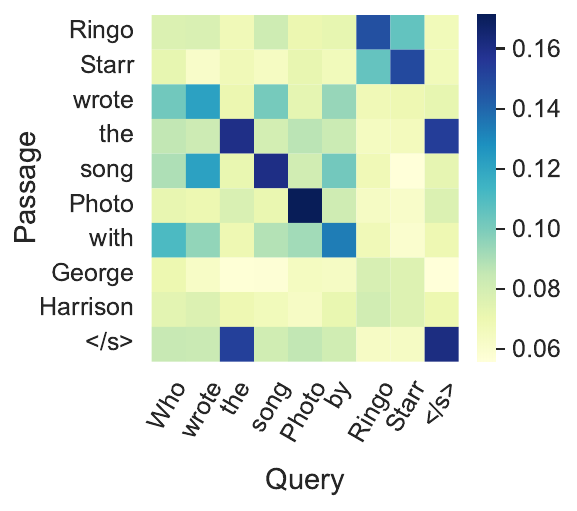}
  \label{fig:case_alignment-mvdr}}
  \hfill
  \subfigure[\ac{GR} query-to-document direction]{\includegraphics[height=0.43\linewidth]{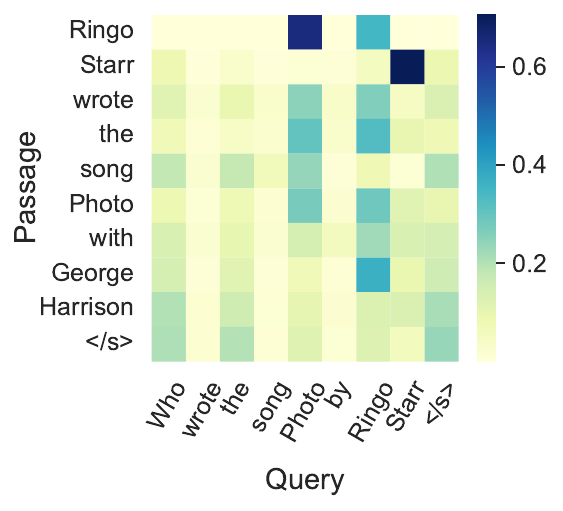}
  \label{fig:case_alignment-gr}}
  \caption{An example of alignment matrix in \ac{MVDR} and \ac{GR}.}
  \label{fig:case_alignment}
\end{figure}

We chose a specific case from the dataset NQ320k to show what the alignment matrix looks like in Figure~\ref{fig:case_alignment}.
Since the document is too long for demonstration, we simplify and extract a sub-sentence containing the answer to the query.
In Figure~\ref{fig:case_alignment-mvdr}, we have observed a pronounced phenomenon of exact matches in \ac{MVDR}.
The song name and people's names are completely matched with high scores.
In Figure~\ref{fig:case_alignment-gr}, this phenomenon is less obvious, but each document token has more attention on the people's name and song name.

\subsection{Upshot}

We verify the existence of ``exact term match,'' a specific alignment scenario, in both paradigms.
We also show the superiority of the alignment direction in \ac{MVDR}.
The improved document encoding and the low-rank nature of the alignment matrix are validated.

\section{Limitations}
We have examined the training target of \ac{GR} and have connected it with \ac{MVDR}, but we have not discussed whether relevance computing can be generalized to the generative style inference.
We have not considered the multi-layer interactions in the cross-attention between query and document for simplicity.

Our framework does not discuss how query-generation augmentation reduces the discrepancy~\citep{Zhuang2022BridgingTG}.
We aim to study how different architectures and identifier designs will affect the alignment and generalization during inference in future works.

\section{Conclusion}
In this paper, we have offered new insights into \ac{GR} from the perspective of \ac{MVDR} that both paradigms share the same frameworks for measuring the relevance between a query and a document.
Both paradigms compute relevance as a sum of products of query and document vectors and an alignment matrix.
We have explored how \ac{GR} applies this framework and differs from \ac{MVDR}.
We have shown that \ac{GR} has simpler document encoding and an alignment strategy with different sparsity and direction.
They also share a low-rank property and can be decomposed into query and document components.
We have conducted extensive experiments to verify our conclusions and found that both methods have commonalities of term matching in the alignment.
We also found that query-to-document alignment direction has better performance than document-to-query.

Based on our findings, practitioners in the field may consider leveraging the shared frameworks highlighted in this study to understand and develop new \ac{GR} methods, and pay more attention to the classic term matching problem underlying \ac{GR} models.

As to future work, we will continue to study how multi-layer attention may affect the framework.
The difference in the generalization properties for new documents between \ac{DR} and \ac{GR}~\citep{nguyen-2023-generative,mehta-etal-2023-dsi,dynamic-retriever2023,cont-learning-gr2023cikm} base on our framework is also an important aspect deserving further investigation.
We will continue to discover new relations in the \ac{GR} paradigm and provide more insights into the methodology.

\begin{acks}
    This research was (partially) funded by
    the Natural Science Foundation of China (62102234, 62372275, 62272274, 62202271, T2293773, 62072279),
    the National Key R\&D Program of China with grant No.2022YFC3303004,
    the Natural Science Foundation of Shandong Province (ZR2021QF129),
    the Hybrid Intelligence Center, a 10-year program funded by the Dutch Ministry of Education, Culture and Science through the Netherlands Organisation for Scientific Research, \url{https://hybrid-intelligence-centre.nl}, 
project LESSEN with project number NWA.1389.20.183 of the research program NWA ORC 2020/21, which is (partly) financed by the Dutch Research Council (NWO),
and
    the FINDHR (Fairness and Intersectional Non-Discrimination in Human Recommendation) project that received funding from the European Union’s Horizon Europe research and innovation program under grant agreement No 101070212.

    All content represents the opinion of the authors, which is not necessarily shared or endorsed by their respective employers and/or sponsors.
\end{acks}

\appendix

\section{Further explanation of the relevance score in \ac{GR}}
\label{app:sim_score_in_gr}
Further insights into the reason for using $\sum_{i\in[M]} \ve_{d_i}^\top \vh_i$ as the relevance can be elaborated as follows.
Suppose we treat all other token embeddings $\ve_k$, where $k\neq d_i$, as fixed with respect to $\ve_{d_i}$, the loss can be expressed as:
\begin{align}
\mc{L}_i(d, q) &= -\ve_{d_i}^\top \vh_i + \log\left(\exp\ve_{d_i}^\top \vh_i + \sum_{v\neq d_i}\exp\ve_v^\top \vh_i\right) \\
&= -\ve_{d_i}^\top \vh_i + \log\left(\exp\ve_{d_i}^\top \vh_i + C\right).
\end{align}
Given that $C\gg \exp{\ve_{d_i}^\top \vh_i}$ at the early stage of the training, we can consider the last term $\log (\exp{\ve_{d_i}^\top \vh_i}+C)$ as a constant w.r.t. $\ve_{d_i}^\top\vh_i$. Hence,
\begin{equation}
\log p(d_i\mid d_{i-1},\ldots,d_0,q)=\mc{L}_i(d, q) \propto \ve_{d_i}^\top \vh_i,
\end{equation}
and further we have
\begin{equation}
\log p(d\mid q)=\mc{L}(d, q)=\sum_i\mc{L}_i(d, q) \propto \sum_{i\in[M]} \ve_{d_i}^\top \vh_i.
\end{equation}
In summary, the dot-product can effectively indicate the relevance between document $d$ and query $q$.

\section{Proof of Lemma~\ref{lemma:gr_alignment_low_rank}}
\label{app:gr_alignment_low_rank}

\begin{proof}
We define $\mR_d=\mD-\vec{1}\vec{d}^\top$, where $\vec{d}=\arg \min \|\mD-\vec{1}\vec{d}^\top\|$.
$\mR_q$ is defined similarly for $\mQ$.
Then we have
\begin{equation*}
    \mD^\top \mW\mQ=(\vd^\top\mW\vq+\mR_d\mW\vq)\vec{1}^\top + (\vec{1}\vd^\top\mW\mR_q^\top+\mR_d\mW\mR_q^\top),
\end{equation*}
therefore, $\mA=\operatorname{softmax}(\mR_d\mW\mR_q^\top+\vec{1}\vec{r}^\top)$.

Let $\mR=\vec{1}\vec{r}^\top$, according to Lemma A.3 in~\citep{dong2021attention}, we have $\|\mA-\mR\|_{1,\infty}\le 4\gamma^2\|\mR_d\mW\mR_q^\top\|_1\le 4\gamma C\|\mW\|_1$,
where $C$ depends on $\mR_d$ and $\mR_q$.
\end{proof}

\section{Proof of Lemma~\ref{lemma:gr_alignment_kernel}}
\label{app:gr_alignment_kernel}
\begin{proof}
Given that $\mA_{ij}=\exp(\vd_i^\top\vq_j)/p_{ij}$, where $p_{ij}$ is the normalization, we can kernelize $\exp(\cdot)$ term as $\phi(\vd_i)^\top\phi(\vq_j)$, and
\begin{align*}
    \operatorname{rel}(d, q)&=\sum_{i,j}\operatorname{tr}(\vd_i^\top\vq_j\phi(\vd_i)^\top\phi(\vq_j))/p_{ij}\\
    &=\sum_{i,j}\frac{1}{p_{ij}}\operatorname{tr}\left(\mat{F}(\vd_i)^\top \mat{F}(\vq_j)\right).\qedhere
\end{align*}
\end{proof}

\section{T5 variants of ColBERT and SEAL}
\label{app:detail_of_T5_variants}
The official ColBERTv1 uses pair-wise contrastive loss for training, which is inefficient.
Therefore, we use in-batch negatives and compute all negative samples simultaneously in each iteration.
We set the batch size to 256 and train 5 epochs.
Using in-batch negatives will greatly save training time without decreasing the performance much.
We summarize our comparison results in Table~\ref{table:t5_reproduce_check_on_nq}.
We also compare the end-to-end performance of T5-ColBERT and T5-SEAL with other commonly used baselines in Table~\ref{table:performance}.

\begin{table}[ht]
\caption{Comparison of our in-batch negative variant of T5-ColBERT and ColBERT with official ColBERTv1~\citep{khattabColBERTEfficientEffective2020}\protect\footnotemark{} on NQ320k~\citep{sunLearningTokenizeGenerative2023}.
``ib neg.'' means in-batch negative.
``R'' denotes Recall, and ``M'' denotes MRR.}
\label{table:t5_reproduce_check_on_nq}
\begin{tabular}{@{}l cccc @{}}
    \toprule
    \textbf{Model} & ib neg. & R@1 & R@10 & M@10 \\
    \midrule
    BM25 & - & 37.6 & 69.5 & 47.8 \\
    Sentence-T5~\citep{sunLearningTokenizeGenerative2023} & - & 53.6 & 83.0 & - \\
    \midrule
    official ColBERTv1~\citep{khattabColBERTEfficientEffective2020} & $\times$ & 57.5 & 84.3 & 66.8 \\
    ColBERT & \checkmark & 58.4 & 86.6 & 68.0 \\
    T5-ColBERT & \checkmark & 61.1 & 88.4 & 70.7 \\
    \bottomrule
\end{tabular}    
\end{table}
\footnotetext{\url{https://github.com/stanford-futuredata/ColBERT/tree/colbertv1}}

\begin{table}[ht]
\caption{End-to-end performance comparison on NQ320k and MS MARCO dataset.
The results of baselines are from~\citep{sunLearningTokenizeGenerative2023,wangNeuralCorpusIndexer2023,zhouUltronUltimateRetriever2022,SemanticEnhancedDSI2023,Zhang2023TermSetsCB}.
``R'' denotes Recall, and ``M'' denotes MRR.}
\label{table:performance}
\begin{tabular}{@{}l cccccc@{}}
    \toprule
     & \multicolumn{3}{c}{\textbf{NQ320K}} & \multicolumn{3}{c}{\textbf{MS MARCO}} \\
    \cmidrule(lr){2-4}    \cmidrule(lr){5-7}
    \textbf{Model} & R@1 & R@10 & M@10 & R@1 & R@10 & M@10 \\
    \midrule
    BM25~\citep{bm25}    & 37.6 & 69.5 & 47.8    & 38.4 & 67.7 & 47.6 \\
    DPR~\citep{karpukhinDensePassageRetrieval2020}     & 50.2 & 77.7 & -       & 49.1 &  76.4 & - \\
    S-T5~\citep{sentence-t5-2022}    & 53.6 & 83.0 & -       & 41.8 & 75.4 & - \\
T5-ColBERT  & 61.1 & 88.4 & 70.7    & 44.2 & 77.8 & 55.3 \\
    \midrule
    DSI~\citep{tayTransformerMemoryDifferentiable2022a}     & 55.2 & 67.4 & -       & 25.7 & 53.8 & 33.9 \\
    T5-SEAL & 44.7 & 75.5 & 55.0    & 24.1 & 57.1 & 34.0 \\
    \bottomrule
\end{tabular}    
\end{table}

\newpage
\bibliographystyle{ACM-Reference-Format}
\balance
\bibliography{references}

\end{document}